\documentclass[aps,amsfonts,preprint]{revtex4}

\begin{document}
\title{ General Unitary TFD Formulation for Superstrings }

\author{M. C. B. Abdalla{\footnote{daniel@ift.unesp.br}}, 
 A. L. Gadelha {\footnote{gadelha@ift.unesp}}
and Daniel L. Nedel{\footnote{daniel@ift.unesp}}}

\affiliation{Instituto de F\'{\i}sica Te\'{o}rica, Unesp, Pamplona 145,
S\~{a}o Paulo, SP, 01405-900, Brazil }

\begin{abstract}
A generalization of the Thermo Field Dynamics (TFD) for
fermionic degrees of freedom is proposed. Such a generalization
follows a previous one where the SU(1,1) thermal group was used to
obtain the closed bosonic string at finite temperature. The SU(2)
thermal group is introduced to construct a general thermal
Bogoliubov transformation to get the type IIB
superstring at finite temperature.
\end{abstract}

\maketitle









Thermo Field Dynamics (TFD) \cite{tu} is a real time approach to
deal with systems at finite temperature. Its basic elements are
the doubling of degrees of freedom of the system under study and a
(thermal) Bogoliubov transformation for entangle such a duplicated
degrees. With the doubling, one obtains an enlarged Hilbert space
composed by the original and an auxiliary space, identical to the
first and related to the so called tilde system. The enlarged
Hilbert space is denoted by a hat and is given by
$\widehat{\mathcal{H}}=\mathcal{H}\otimes\widetilde{\mathcal{H}}$.
The original and tilde systems are related by a mapping called
tilde conjugation rules. These rules are related with the
Tomita-Takesaki modular operator of the statistical mechanics
algebraic approach \cite{oji,land}.

The thermal Bogoliubov transformation is obtained using a generator
in such a way that, in a finite volume limit, the transformation is
unitary and preserves the tilde conjugation rules.
The thermal effects arise from the vacuum correlation introduced by
the transformation over the enlarged system vacuum.

The construction above outlined was the first one for the TFD.
However, one can find a set of generator that maintains the thermal
nature of the transformation. The set of generators is shown to be
a linear combination of operators that forms an oscillator
representation of $SU(1,1)$ group for bosons and $SU(2)$ for
fermions \cite{ume,chume}. These sets for fermionic and bosonic
systems can be constructed in two different manner, providing two
possible generalizations of the TFD approach.
In one case the tilde conjugation rules are preserved but the
transformation, in a finite volume limit, is non-unitary. This
construction was largely developed as one can see, for example in
Ref. \cite{ume}. In the second case the transformation is unitary,
in a finite volume limit, but the tilde conjugation rules are not
preserved.

The unitary case was applied to bosonic string and $D_{p}$-brane
in Ref. \cite{agv2}. In a recent paper \cite{gfu} the general
unitary $SU(1,1)$ TFD formulation was proposed for bosonic
systems. Such a formulation considers a transformed
Tomita-Takesaki modular operator, once that it does not commute with
the general generator of the thermal transformation \cite{elmume}.
As a consequence, the tilde conjugation rules were redefined, in
the transformed space, and called breve conjugation rules. The
generalized $SU(1,1)$ thermal vacuum is invariant under breve
conjugation.

The objective of this work is to extend such a unitary TFD
formulation to include fermionic degrees of freedom, using as
target system the type IIB Green-Schwarz superstring. Here, a
TFD construction taking into account explicitly the level-matching
condition is applied \cite{nos,nos2}.
We consider the light-cone coordinates
$X^\pm = \frac{1}{\sqrt{2}}\left(X^9 \pm X^0\right)$ and
we write the remaining 8 components of the spinors (after Kappa
symmetry fixing) as $S^a$, $\bar{S}^a$, composing the ${\bf 8_{s}}$
representation of $SO(8)$. The chiral representation of $SO(8)$ gamma
matrices is used.

The solutions of the equation of motion for the system can be
expanded in string modes. After the quantization we choose the
oscillator description of such a modes
%
in order to deal with the standard commutation and
anti-commutation relations of the harmonic oscillator. The
left-moving bosonic and fermionic oscillators are denoted by
$a_{n}^{I}$, $a_n^{\dagger\: I}$ and $S_{n}^{a}$,
$S_{n}^{\dagger\: a}$. The right-moving modes  will be denoted
by a ``bar'' over the operators.
In general the vacuum of the fermionic zero modes
are chosen to be a set of states denoted by
 ${\left|I\right\rangle,\left|\dot{a}\right\rangle }$, satisfying:
$ S_0^a\left|I\right\rangle = \gamma^I_{a\dot{a}}
\left|\dot{a}\right\rangle$ and $S_0^a \left|\dot{a}\right\rangle=
\gamma^I_{a\dot{a}} \left|I\right\rangle$ \cite{GS}.
In order to apply the TFD
algorithm it is necessary to use a different fermionic vacuum than
the usual one. The vacuum is chosen to be a state
$\left|0,p^+\right\rangle$ such that
\begin{eqnarray}
S^{a}\left|0,p^+\right\rangle&=&0,
\nonumber \\
S_{n}^{a}\left|0,p^+\right\rangle
&=&\bar{S}_{n}^{a}\left|0,p^+\right\rangle=0, \qquad n>0,
\nonumber \\
a_{n}^{I}\left|0,p^+\right\rangle
&=&\bar{a}_{n}^I\left|0,p^+\right\rangle=0, \qquad n>0,
\label{vac}
\end{eqnarray}
with the following definition for the fermionic zero modes
creation-annihilation operators:
\begin{eqnarray}
S^a &=&\frac{1}{\sqrt{2}}\left(S_{0}^{a}+i\bar{S}_{0}^{b}\right),
\qquad
S^{\dagger\:a}=\frac{1}{\sqrt {2}}\left(S_{0}^{a}-
i\bar{S}_{0}^{a}\right),
\end{eqnarray}
which satisfies
\begin{equation}
\left\{ S^{a}, S^{\dagger\:b}\right\}=\delta^{ab},\qquad
\left\{ S^{a}, S^{b}\right\}=\left\{ S^{\dagger\:a},
S^{\dagger\:b}\right\}=0.
\end{equation}

The light-cone Hamiltonian is calculated in a standard way and it
is written as
\begin{equation}
p_{+} H= \frac{p^{i}p^{i}}{2}+
\sum_{n=1}n\left(a_{n}^{\dagger\:I}a_{n}^{I}
+\bar{a}_{n}^{\dagger\:I}\bar{a}_{n}+
S_{n}^{\dagger\:a}S_{n}^{a} +
\bar{S}_{n}^{\dagger\:a}\bar{S}_{n}^{a}\right).
\label{aga}
\end{equation}

The 32 supersymmetries are divided in a set of  $16$ kinematical
supercharges, that belong to ${\bf 8_{s}}$ of $SO(8)$, and $16$
dynamical supercharges, that transform the fields of the same
supermultiplet and belong to ${\bf 8_{c}}$ of $SO(8)$. The kinematical
and dynamical
supercharges can be written as $Q_{a}^{\pm}=Q_{a} \pm i\bar{Q}_{a}$
and $Q_{\dot a}^{\pm}=Q_{\dot a} \pm i\bar{Q}_{\dot a}$,
respectively, where
\begin{eqnarray}
\sqrt{p^{+}}Q_{a}&=&S^{a}_{0},
\nonumber
\\
\sqrt{p^+}Q_{\dot a}&=&
P_{0}^I\left(\gamma^I S_{0}\right)_{\dot a}
+ \sum_{n=1}\left[\sqrt{2\omega_{n}}\left(a_{n}^{\dagger\:I}
\gamma^{I}S_{n}
+ a_{n}^{I}\gamma^{I}S_{n}^{\dagger}\right)_{\dot a}\right],
\end{eqnarray}
and $\bar{Q}_{\dot{a}}$  can be obtained from $Q_{\dot{a}}$
replacing ``bar'' variables by non-bar variables and $i$ by $-i$.
It can be seen that these superchages annihilate the vacuum
$(\ref{vac})$.

To construct the physical Fock space it is necessary to fix the
residual gauge symmetry generated by the world sheet momentum $P$.
This gauge fixing improves the level matching condition on a physical
state $\left |\Phi\right\rangle$:
\begin{equation}
P\left |\Phi\right\rangle= \sum_{n=1}^{\infty }
n\left( N_{n}^{B}+N_{n}^{F}-\bar{N}_{n}^{B}
-\bar{N}_{n}^{F}\right)\left |\Phi\right\rangle=0,
\label{lmc}
\end{equation}
where the $N^{B}$, $N^{F}$, are the bosons and fermions number operators
for left-moving modes and the ``bar'' ones the same for the right-moving
modes.

Let us now apply the TFD approach to construct the thermal Fock
space for the type IIB superstring.
We have first to duplicate the degrees of freedom. To this
end we consider a copy of the original Hilbert space, denoted by
$\widetilde{H}$. The tilde Hilbert space is built with the set
of oscillators operators, $\widetilde{a}_{0}$, $\widetilde{S}^{a}$,
$\widetilde{a}_{n}^{I}$, $\widetilde{\bar{a}}_{n}^{I}$,
$\widetilde{S}_{n}^{a}$, $\widetilde{\bar{S}}_{n}^{a}$ that have the same
(anti-) commutation properties of the original ones and the operators of
the two systems (anti-) commute among themselves.

We can now construct the thermal system. This is achieved by implementing
a thermal Bogoliubov transformation in the total (enlarged) Hilbert space.
The transformation generator considered in our generalization is given by
\begin{equation}
G=G^{B}+G^{F},
\label{gen}
\end{equation}
for
\begin{equation}
G^{B}=\sum_{n=1} \left(G_{n}^{B} + \bar{G}_n^{B}\right),
\qquad
G^{F}=G_{0}^{F}+ \sum_{n=1} \left(G_{n}^{F}
+ \bar{G}_{n}^{F}\right),
\label{genf}
\end{equation}
where
\begin{eqnarray}
G_{0}^{F} &=&\gamma_{1_{0}}\widetilde{S}^{\dagger}
\cdot S^{\dagger }
-\gamma_{2_{0}} S\cdot \widetilde{S}
+\gamma
_{3_{0}}\left( S^{\dagger }\cdot S -
\widetilde{S} \cdot \widetilde{S}^{\dagger }\right) ,
\label{gef0}
\\
G^{B}&=& \sum_{n=1}\left[\lambda _{1_{n}}\widetilde{a}_{n}^{\dagger}
\cdot a^{\dagger }_{n}
-\lambda_{2_{n}} a_{n}\cdot \widetilde{a}_{n}
+\lambda
_{3_{n}}\left( a^{\dagger }_{n}\cdot a_{n} +
\widetilde{a}_{n} \cdot \widetilde{a}^{\dagger }_{n}\right)\right] ,
\label{geb}
\\
G^{F}&=& \sum_{n=1}\left[\gamma_{1_{n}}\widetilde{S}_{n}^{\dagger}
\cdot S^{\dagger }_{n}
-\gamma_{2_{n}} S_{n}\cdot \widetilde{S}_{n}
+\gamma
_{3_{n}}\left( S^{\dagger }_{n}\cdot S_{n} -
\widetilde{S}_{n} \cdot \widetilde{S}^{\dagger }_{n}\right)\right] ,
\label{gef}
\end{eqnarray}
with the $\lambda$ and $\gamma$ coefficients given by
\begin{eqnarray}
\lambda_{1_{n}} &=&\theta_{1_{n}}^{B}-i\theta
_{2_{n}}^{B}, \qquad \lambda_{2_{n}} =-\lambda
_{1_{n}}^{*},\qquad \lambda_{3_{n}} =\theta
_{3_{n}}^{B},
\nonumber
\\
\gamma _{1_{n}}& =& \theta _{1_{n}}^{F}-i\theta
_{2_{n}}^{F}, \qquad \gamma_{2_{n}} =-\gamma
_{1_{n}}^{*},\qquad \gamma_{3_{n}} =\theta
_{3_{n}}^{F},
\nonumber
\\
\gamma _{1_{0}}& =& \theta _{1_{0}}^{F}-i\theta
_{2_{0}}^{F}, \qquad \gamma_{2_{0}} =-\gamma
_{1_{0}}^{*},\qquad \gamma_{3_{0}} =\theta
_{3_{0}}^{F}.
\label{gammadef}
\end{eqnarray}
The labels $B$ and $F$ specify fermions and bosons,
the dots represent the inner products and $\theta$, $\bar{\theta}$
are real parameters. In the thermal equilibrium they are related to
the Bose-Einstein and Fermi-Dirac distributions of the oscillator
$n$ as we will see. The expression (\ref{geb})-(\ref{gammadef})
are the same for ``bar'' sector, but replacing operators and parameters
by ``bar'' ones.

The bosonic and fermionic transformed operators can be obtained
as follows: Consider an oscillator-like operator, $A$, that can
be commuting or anti-commuting. For this operator one have
\begin{eqnarray}
\left(
\begin{array}{c}
A^{i}_{n}(\theta) \\
\breve{A}^{i \dagger}_{n}(\theta)
\end{array}
\right) &=&e^{-iG}\left(
\begin{array}{c}
A^{i}_{n} \\
\widetilde{A}^{i\dagger }_{n}
\end{array}
\right) e^{iG}={\mathbb B}_{n}\left(
\begin{array}{c}
A^{i}_{n} \\
\widetilde{A}^{i \dagger }_{n}
\end{array}
\right) , \label{tbt}
\\
\left(
\begin{array}{cc}
A^{i \dagger}_{n}(\theta) & -\sigma\breve{A}^{i}_{n}(\theta)
\end{array}
\right) &=&\left(
\begin{array}{cc}
A^{i \dagger }_{n} & -\sigma\widetilde{A}^{i}_{n}
\end{array}
\right) {\mathbb B}^{-1}_{n}, \label{tbti}
\end{eqnarray}
where the matrix transformation is given by
\begin{eqnarray}
{\mathbb B}_{n}=\left(
\begin{array}{cc}
u_{n} & v_{n} \\
\sigma v^{*}_{n} & u^{*}_{n}
\end{array}
\right) ,\qquad |u_{n}|^{2}-\sigma |v_{n}|^{2}=1, \label{tbm}
\end{eqnarray}
with $\sigma=1$ for bosons and $\sigma=-1$ for fermions.
The matrix elements for fermions are
\begin{equation}
u_{n}\equiv U_{n}^{F}=\cosh \left( i\Gamma_{n} \right)
+\frac{\gamma_{3_{n}}}{\Gamma_{n} }
\sinh\left(i\Gamma_{n} \right),
\qquad
v_{n}\equiv V_{n}^{F}=-\frac{\gamma _{1_{n}}}{\Gamma_{n} }\mbox{sinh}\left(
i\Gamma_{n} \right) , \label{uvexpF}
\end{equation}
and $\Gamma_{n}$ is defined by the following relation
\begin{equation}
\Gamma ^{2}_{n}=-\gamma _{1_{n}}\gamma _{2_{n}}+\gamma
_{3_{n}}^{2}.
\label{Gadef}
\end{equation}
For bosons one has
\begin{equation}
u_{n}\equiv U_{n}^{B}=\cosh \left( i\Lambda_{n} \right) +\frac{\lambda
_{3_{n}}}{\Lambda_{n} } \sinh\left(i\Lambda_{n} \right),
\qquad
v_{n}\equiv V_{n}^{B}=\frac{\lambda _{1_{n}}}{\Lambda_{n} }\mbox{sinh}\left(
i\Lambda_{n} \right) , \label{uvexpB}
\end{equation}
and $\Lambda_{n}$ is defined by the following relation
\begin{equation}
\Lambda ^{2}_{n}=\lambda _{1_{n}}\lambda _{2_{n}}+\lambda
_{3_{n}}^{2}.
\label{Ladef}
\end{equation}

The expression (\ref{tbt})-(\ref{Ladef}) for the ``bar'' sector can be
obtained replacing the operators and parameters by ``bar'' ones.
The new conjugation rules in the transformed space, called breve,
can be presented as follows: consider any operators $A\left(\theta \right)$,
$B\left(\theta \right)$ in the transformed space. These operators can be
fermionic or bosonic. Associated with these operators one has their breve
counterparts $\breve{A}\left(\theta \right)$,
$\breve{B}\left(\theta \right)$. The map between both spaces
is given by the following breve conjugation rules
\begin{eqnarray}
\left[ A\left(\theta \right)B\left(\theta \right)\right] \breve{^{}}
&=&\breve{A} \left(\theta \right)\breve{B}\left(\theta \right),
\label{breve1}
\\
\left[ c_{1}A\left(\theta \right)+ c_{2}B\left(\theta \right)\right]
\breve{^{}} &=&\left[
c_{1}^{*}\breve{A}\left(\theta\right)+c_{2}^{*}\breve{B}\left(\theta
\right)\right] ,
\label{breve2}
\\
\left[ A^{\dagger }\left(\theta \right)\right] \breve{^{}}
&=&\breve{A}^{\dagger }\left(\theta \right),
\label{breve3}
\\
\left[ \breve{A}\left(\theta\right)\right] \breve{^{}}
&=&\sigma A\left(\theta \right),
\label{breve4}
\\
\left[ \left| 0\left( \theta \right) \right\rangle
\right] \breve{^{}}
&=&\left| 0\left( \theta \right) \right\rangle ,
\label{tvbinv}
\\
\left[\left\langle 0\left(\theta\right)\right|\right]
\breve{^{}}
&=&\left\langle 0\left( \theta \right) \right|,
\label{tbbinv}
\end{eqnarray}
with $c_{1}$, $c_{2} \in \mathbb{C}$.
The expressions (\ref{tvbinv}) and (\ref{tbbinv}) inform that
the thermal vacuum obtained from the thermal
Bogoliubov transformation with the generator presented at (\ref{gen})
and given by
\begin{eqnarray}
\left |0\left(\theta\right)\right\rangle \! &=& \! e^{-i{G}}\left.
\left|0\right\rangle \!\right\rangle \nonumber
\\
\!&=&\!
\left(U_{0}^{F}\right)^{8}
e^{-\frac{V_{0}^{F}}{U_{0}^{F}}S_{0}^{\dagger} \cdot
{\widetilde S}_{0}^{\dagger}}
\prod_{n=1}\!\left[\left(
\frac{U_{n}^{F}}{U_{n}^{B}}\right)^{\!\!\!8}\left( \frac{\bar
{U}_{n}^{F}}{\bar{U}_{n}^{B}}\right)^{\!\!\!8}
e^{-\frac{V_{n}^{B}}{U_{n}^{B}}a_{n}^{\dagger}\cdot
\widetilde{a}_{n}^{\dagger}
- \frac{\bar{V}_{n}^{B}}{\bar{U}_{n}^{B}}
\bar{a}_{n}^{\dagger}\cdot \widetilde{\bar{a}}_{n}^{\dagger}
-\frac{V_{n}^{F}}{U_{n}^{F}}
S_{n}^{\dagger}\cdot \widetilde{S}_{n}^{\dagger}
-\frac{\bar{V}_{n}^{F}}{\bar{U}_{n}^{F}}\bar{S}_{n}^{\dagger}
\cdot \widetilde {\bar{S}}_{n}^{\dagger}}\right]
\!\!\left.\left|0\right\rangle\!\right\rangle \!\!, \nonumber \\
\label{tva}
\end{eqnarray}
is invariant under breve conjugation rules.
The thermal Fock space is constructed from the above vacuum
by applying the thermal creation operators.
As the Bogoliubov transformation is canonical, the thermal
operators obey the same (anti-) commutators relations as the
operators at $T=0$.

According to the TFD formulation the hamiltonian of the total
(duplicated) system in constructed in order to keep untouched the
original dynamics of the superstring. This operator can be written
as
\begin{equation}
{\widehat H} = H - \widetilde{H},
\label{hath}
\end{equation}
in such a way that ${\widehat H}$ plays the
r\^{o}le of the hamiltonian generating the temporal translation in
the thermal Fock space. Using the commutation relations we can
prove that the Heisenberg equations are satisfied replacing $H$ and
${\widetilde H}$ by ${\widehat H}$.


The TFD approach will be now used to compute thermodynamical
quantities by evaluating operators from the original system in the
thermal Fock space. It was appointed out by Polchinski \cite{pol}
that, in the one-string sector, the torus path integral
computation of the free energy coincides with what we would obtain
by adding the contributions from different states of the spectrum
to the free energy. In a recent work \cite{nos3} it was shown how
the closed string thermal torus can be viewed in TFD approach.

The free energy is obtained from the knowledge of the thermal
energy and entropy operators. The energy operator is such that the
level matching condition must be implemented. In the TFD approach
it can be done by considering the shifted hamiltonian:
\begin{equation}
H_{s}=H+\frac{ 2\pi i\lambda}{\beta}P,
\end{equation}
where $H$, the original hamiltonian, is defined at (\ref{aga}),
$\lambda$ is a lagrange multiplier and $P$ is given at (\ref{lmc}).
The dependence of the thermal vacuum on the lagrange multiplier comes
from the Bogoliubov transformation parameters. This is achieved defining
first the free energy like potential
\begin{equation}
{\cal F} = {\cal E} - \frac{1}{\beta}{\cal S},
\label{freed}
\end{equation}
with
\begin{equation}
{\cal E} \equiv \left\langle 0 \left(\theta\right)
\right| H_{s}\left|0\left(\theta\right)\right\rangle,
\qquad
{\cal S} \equiv \left\langle 0\left(\theta\right)
\right| K \left| 0\left( \theta \right) \right\rangle,
\end{equation}
where ${\cal E}$ is related with thermal energy of the system, and
${\cal S}$ with its entropy obtained from the entropy operator,
$K$, defined as follows
\begin{equation}
K=K^{B}+K^{F},
\label{kab}
\end{equation}
where the bosonic entropy operator is given by
\begin{eqnarray}
K^{B}&=&
-\sum_{n=1}\left\{a_{n}^{\dagger }\cdot a_{n}
\ln\left(\frac{\lambda _{1_{n}}\lambda_{2_{n}}}
{\Lambda^2_{n}}\sinh^{2}\left(i\Lambda_{n}\right)\right)
-a_{n}\cdot a_{n}^{\dagger }
\ln\left(1+\frac{\lambda _{1_{n}}\lambda_{2_{n}}}
{\Lambda^2_{n}}\sinh^{2}\left(i\Lambda_{n}\right)\right)\right\}
\nonumber
\\
&&-\sum_{n=1}\left\{\bar{a}_{n}^{\dagger }\cdot \bar{a}_{n}
\ln\left(\frac{\bar{\lambda}_{1_{n}}\bar{\lambda}_{2_{n}}}
{\bar{\Lambda}^2_{n}}
\sinh^{2}\left(i \bar{\Lambda}_{n}\right)\right)
-\bar{a}_{n}\cdot \bar{a}_{n}^{\dagger }
\ln\left(1+\frac{\bar{\lambda}_{1_{n}}\bar{\lambda}_{2_{n}}}
{\bar{\Lambda}^2_{n}}
\sinh^{2}\left(i \bar{\Lambda}_{n}\right)\right)\right\},
\nonumber
\end{eqnarray}
and the fermionic one by
\begin{eqnarray}
K^{F}&=&
-\left\{S_{0}^{\dagger }\cdot S_{0}
\ln\left(\frac{\gamma _{1_{0}}\gamma_{2_{0}}}
{\Gamma^2_{0}}\sinh^{2}\left(i\Gamma_{0}\right)\right)+
S_{0}\cdot S_{0}^{\dagger }
\ln\left(1-\frac{\gamma _{1_{0}}\gamma_{2_{0}}}
{\Gamma^2_{0}}\sinh^{2}\left(i\Gamma_{0}\right)\right)\right\}
\nonumber
\\
&&-\sum_{n=1}\left\{S_{n}^{\dagger }\cdot S_{n}
\ln\left(\frac{\gamma _{1_{n}}\gamma_{2_{n}}}
{\Gamma^2_{n}}\sinh^{2}\left(i\Gamma_{n}\right)\right)
+S_{n}\cdot S_{n}^{\dagger }
\ln\left(1-\frac{\gamma _{1_{n}}\gamma_{2_{n}}}
{\Gamma^2_{n}}\sinh^{2}\left(i\Gamma_{n}\right)
 \right)\right\}
\nonumber
\\
&&-\sum_{n=1}\left\{\bar{S}_{n}^{\dagger }\cdot
\bar{S}_{n}\ln\left(\frac{\bar{\gamma}_{1_{n}}\bar{\gamma}_{2_{n}}}
{\bar{\Gamma}^2_{n}}
\sinh^{2}\left(i \bar{\Gamma}_{n}\right)\right)
+\bar{S}_{n}\cdot \bar{S}_{n}^{\dagger }
\ln\left(1-\frac{\bar{\gamma}_{1_{n}}\bar{\gamma}_{2_{n}}}
{\bar{\Gamma}^2_{n}}
\sinh^{2}\left(i \bar{\Gamma}_{n}\right)\right)\right\}.
\nonumber
\end{eqnarray}

Performing the above expectation values and minimizing
the free energy like potential with respect to the parameters
it will be a minimum for
\begin{eqnarray}
|V_{0}^{F}|^2
&=&
\frac{\gamma _{1_{0}}\gamma_{2_{0}}}
{\Gamma^2_{0}}\sinh^{2}\left(i\Gamma_{0}\right)=\frac{1}{2},
\nonumber
\\
|{V}_{n}^{F}|^{2}
&=&
\frac{\gamma _{1_{n}}\gamma_{2_{n}}}
{\Gamma^2_{n}}\sinh^{2}\left(i\Gamma_{n}\right)
=\frac{1}{e^{n\left(\frac{\beta}{p+}+i2\pi\lambda\right)}+1},
\qquad
\nonumber
\\
|\bar{V}_{n}^{F}|^{2}
&=&
\frac{\bar{\gamma}_{1_{n}}\bar{\gamma}_{2_{n}}}
{\bar{\Gamma}^2_{n}}
\sinh^{2}\left(i \bar{\Gamma}_{n}\right)
=\frac{1}{e^{n\left(\frac{\beta}{p+}-i2\pi\lambda\right)}+1},
\nonumber
\\
|V_{n}^{B}|^{2}
&=&
\frac{\lambda _{1_{n}}\lambda_{2_{n}}}
{\Lambda^2_{n}}\sinh^{2}\left(i\Lambda_{n}\right)
=\frac{1}{e^{n\left(\frac{\beta}{p^{+}}+i2\pi\lambda\right)}-1},
\qquad
\nonumber
\\
|\bar{V}_{n}^{B}|^{2}
&=&
\frac{\bar{\lambda}_{1_{n}}\bar{\lambda}_{2_{n}}}
{\bar{\Lambda}^2_{n}}
\sinh^{2}\left(i \bar{\Lambda}_{n}\right)=
\frac{1}{e^{n\left(\frac{\beta}{p^{+}}-i2\pi\lambda\right)}-1}.
\label{parmin}
\end{eqnarray}
The above expressions manifest the dependence of the transformation
parameters with the lagrange multiplier $\lambda$, $p^{+}$ and $\beta$,
and fix the thermal vacuum $\left(\ref{tva}\right)$
as those that reproduce the trace over the transverse sector.

Replacing the above results in the free energy like potential
(\ref{freed}), one finds, following \cite{nos,nos2},
\begin{equation}
F=-\frac{l_{s}}{\beta}\int dp^{+}\int_{0}^{1}d\lambda \: e^{\beta p^{+}}
\ln\: 2^{8} e^{-\beta\frac{p^{2}}{2p^{+}}}
\prod_{n={\mathbb Z}}
\left[ \frac{1+e^{-\frac {\beta \omega_{n}}{p+}+ i \lambda k_{n}}}
{1-e^{-\frac {\beta \omega_{n}}{p+}+i \lambda k_{n}}}\right]^{8}.
\label{final}
\end{equation}
This expression is the TFD answer for the type IIB Green-Schwarz
superstring free energy. The above result is the same of that
obtained by using other methods as for example the statistical
mechanics operator approach and functional integration, when the
world sheet is defined on a torus.

The factor $2^8$ comes from the zero modes of the entropy
operator. In fact, it is the degeneracy of the ground state and
corresponds to the contribution of the 256 massless states of the type
IIB superstring to the free energy. With this the TFD
generalization presented here seems to be consistent at least for
the system studied in this work.

We would like to thank D. Z. Marchioro and I. V. Vancea for useful
discussions. M. C. B. A. was partially supported by the CNPq Grant
302019/2003-0, A. L. G. and D. L. N. are supported by a FAPESP
post-doc fellowship.

\end{document}